\documentclass{jfm}

\ifCUPmtlplainloaded \else
  \checkfont{eurm10}
  \iffontfound
    \IfFileExists{upmath.sty}
      {\typeout{^^JFound AMS Euler Roman fonts on the system,
                   using the 'upmath' package.^^J}%
       \usepackage{upmath}}
      {\typeout{^^JFound AMS Euler Roman fonts on the system, but you
                   dont seem to have the}%
       \typeout{'upmath' package installed. JFM.cls can take advantage
                 of these fonts,^^Jif you use 'upmath' package.^^J}%
      }
  \else
  \fi
\fi

\ifCUPmtlplainloaded \else
  \checkfont{msam10}
  \iffontfound
    \IfFileExists{amssymb.sty}
      {\typeout{^^JFound AMS Symbol fonts on the system, using the
                'amssymb' package.^^J}%
       \usepackage{amssymb}%
       \let\le=\leqslant  
       \let\ge=\geqslant  
      }{}
  \fi
\fi

\ifCUPmtlplainloaded \else
  \IfFileExists{amsbsy.sty}
    {\typeout{^^JFound the 'amsbsy' package on the system, using it.^^J}%
     \usepackage{amsbsy}}
    {\providecommand\boldsymbol[1]{\mbox{\boldmath $##1$}}}
\fi

\providecommand\bnabla{\boldsymbol{\nabla}}
\providecommand\bcdot{\boldsymbol{\cdot}}

\newcommand\taub{\boldsymbol{\tau}}
\newcommand\figref[1]{figure~\ref{#1}}

\newcommand\Imag{\mbox{Im}} 

\title[Viscous heating effects in temperature-dependent viscosity
  fluids]{Viscous heating effects in fluids with temperature-dependent
  viscosity: triggering of secondary flows}
\author[A. Costa and G. Macedonio]%
{A.\ns C\ls O\ls S\ls T\ls A$^1$%
\ns \and
G.\ns M\ls A\ls C\ls E\ls D\ls O\ls N\ls I\ls O$^1$}
\affiliation{$^1$Osservatorio Vesuviano - Istituto Nazionale di
  Geofisica e Vulcanologia, Via Diocleziano 328, Naples,
  Italy\\[\affilskip]} 

\pubyear{2005}
\volume{}
\pagerange{}
\date{?? and in revised form ??}
\newcommand\paref[1]{(\ref{#1})}

\usepackage{natbib}
\usepackage[dvips,final]{graphicx}

\begin{document}

\maketitle
\begin{abstract}
Viscous heating can play an important role in the dynamics of fluids
with a strongly temperature-dependent viscosity because of the
coupling between the energy and momentum equations.
The heat generated by viscous friction produces a local increase in
temperature near the tube walls with a consequent decrease of the
viscosity and a strong stratification in the viscosity profile which
can cause a triggering of instabilities and a transition to secondary
flows. The problem of viscous heating in fluids was investigated and
reviewed by \citet{cosmac2003} for its important implications
in the study of magma flows.\\ 
In this paper we present two separate theoretical models: a linear
stability analysis and a direct numerical simulation (DNS) of a plane
channel flow. In particular  DNS shows that, in certain regimes, viscous
heating can trigger and sustain a particular class of secondary
rotational flows which appear organized in coherent structures similar
to roller vortices. This phenomenon can play a very important role in
the dynamics of magma flows and, to our knowledge, it is the first
time that it has been investigated by a direct numerical simulation.   
\end{abstract}
\section{Introduction}
In this paper we show that the effects of viscous heating can play an
important role in the channel flow dynamics of fluids with a strongly
temperature-dependent viscosity such as silicate melts and polymers.  
In fact, in these fluids, viscous friction generates a local increase
in temperature near the channel walls with a consequent viscosity
decrease and often a rise of the flow velocity. This velocity increase
may produce a further growth of the local temperature.
As recently described in \citet{cosmac2003}, above some critical
values of the parameters of the process, this feedback cannot
converge. In this case the one-dimensional laminar solution, valid in
the limit of an infinitely long channel, cannot exist even for low
Reynolds numbers. In channels of finite length, viscous heating
governs the evolution from a Poiseuille regime with a uniform
temperature distribution at the inlet, to a plug flow with a hotter
boundary layer near the walls downstream \citep{pea77,ock79}.     
We will show that when the temperature gradients induced by viscous
heating are relatively large, local instabilities occur and a 
triggering of secondary flows is possible because of viscosity
stratification. 
\par
From previous results \citep[see ][and references
  therein]{cosmac2003}, we know that, in steady state conditions for a
fully developed Poiseuille or Couette flow, there is a critical value
of a  dimensionless ``shear-stress'' parameter 
${\mathcal G}=\beta(\frac{dP}{dx})^2H^4/(k\mu_0)$ (see below for the
symbols used), such that if ${\mathcal G} > {\mathcal G}_{crit}$, then
the system does not admit solution, whereas when ${\mathcal G} <
{\mathcal G}_{crit}$, the  system has two solutions, one of which (the
solution with greater temperature) may be unstable.
For finite length plane channels, \citet{cosmac2003} have shown that
these processes are controlled principally by the P\'eclet number
$Pe$, the  Nahme number $Na$  (also called Brinkman number), and the
non-dimensional flow rate $q$:  
\begin{equation}
Pe=\rho c_p U H/k; \quad Na=\mu_0 U^2\beta/k; \quad 
q=\mu_0Q/(\rho gH^3) 
\end{equation}
with $\rho$ density, $c_p$ specific heat, $U$ mean velocity,
$H$ half channel thickness, $k$ thermal conductivity, $\mu_0$
reference viscosity, $\beta$ rheological parameter (see
equation~(\ref{eq:viscosity_exponential})), $Q$ flow rate per unit
length ($Q=UH$) and $dP/dx$ longitudinal pressure gradient.\\
The characteristic length scales involved are the channel dimensions
$H$ (thickness) and $L$ (length), the mechanical relaxation length
$L_m=UH^2\rho/\mu_0$, and the thermal relaxation length 
$L_t=UH^2\rho  c_p/k$. For magma flows, typically $L_t/L_m \gg 1$ and
the approximation of infinitely long channel (from a thermal point of
view) is not valid. For finite length channels, when viscous heating
is important, starting with uniform temperature  and parabolic
velocity profile at the inlet, the flow evolves gradually to a
plug-like velocity profile with two symmetric peaks in the temperature 
distribution. The more important viscous dissipation effects are, the
more pronounced the temperature peaks are, the lower the length scale
for the development of the plug flow is \citep{ock79,cosmac2003}. 
\par
Because of the typically low thermal conductivities of liquids such as
silicate melts, the temperature field shows a strong transversal
gradient. Flows with layers of different viscosity were
investigated in the past, for their practical interest, and it is
known that they can be unstable depending on their configuration
\citep{yih67,cra69,renjos85,ren87,liren99}. In particular, we find
that when the viscous heating produces a relatively hot less viscous
layer near the wall, there is the formation of spatially periodic
waves and of small vortices near the wall, similar to the waves and
vortices which form in core-annular flows of two fluids with high
viscosity ratio \citep{liren99}. 
\par
In this paper we focus our investigations to the physical regime that
typically characterizes magma flow, with low Reynolds number $Re <
O(10^2)$,  high P\'eclet number $Pe \gg 1$, high Prandtl number $Pr\gg
1$ and low aspect ratio $a_r=H/L \ll 1$ \citep[see
  e.g.\ ][]{wyllis95}.\\
In \S~\ref{ge} we present the governing equations, in \S~\ref{stab} we
analyze the linear stability of the base flow given by a lubrication
approximation, in \S~\ref{numsim} we describe the numerical scheme and
the parameters used for the direct numerical simulation (DNS), then we
discuss the results obtained from DNS and, briefly, few implications
for magma flows.     
\section{Governing equations \label{ge}}
\begin{figure}
\includegraphics[angle=0,width=\hsize]{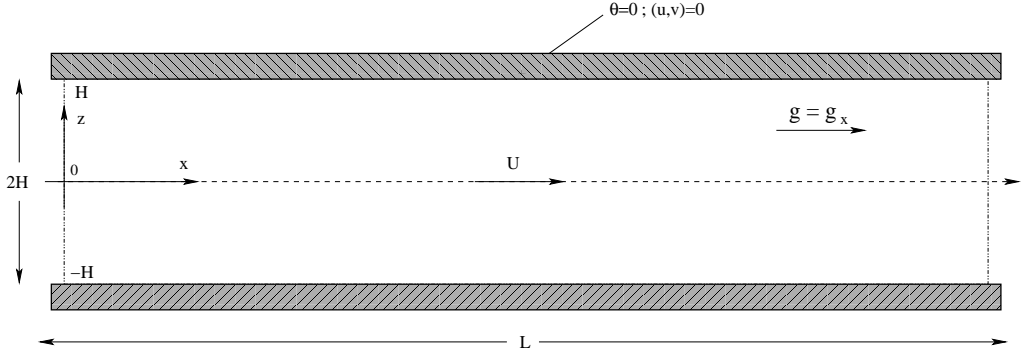}
\caption{Sketch of the studied system: coordinates and channel
  dimensions.} 
\label{fig:schema}
\end{figure}

We consider an incompressible homogeneous fluid with constant
density, specific heat and thermal conductivity. The fluid 
viscosity $\mu$ is temperature-dependent and, although an
Arrhenius-type law of viscosity-temperature dependence relationship is     
more general and adequate to describe, for example, the silicate melt
viscosities, for simplicity in this study we assume the exponential
(Nahme's) approximation:     
\begin{eqnarray}
\mu = \mu_0 \exp[-\beta(T-T_0)]
\label{eq:viscosity_exponential}
\end{eqnarray}
where $T$ is temperature, $\beta$ a rheological factor and $\mu_0$ is
the viscosity value at the reference temperature $T_0$.
Although a strong viscosity-temperature dependence similar to
(\ref{eq:viscosity_exponential}), can be responsible for different
types of magma instabilities, there have been only few studies of them
\citep[see e.g.\ ][]{wyllis95,wyllis98}.\\ 
Here we investigate the two-dimensional flow in a plane channel
between two parallel boundaries of length $L$ separated by a distance
$2H$ (with $H/L\ll 1$) and we restrict our study to a
body-force-driven flow (see \figref{fig:schema}), although it is not
difficult to generalize for pressure-driven flow or up-flow conditions
for which the driving pressure gradient and the gravity act in the
opposite direction, as it occurs for example in magma conduits. \\
In these hypotheses, the fluid dynamics are described by the 
following transport equations for mass, momentum and energy,
respectively: 
\begin{equation}
\bnabla \bcdot {\mathbf v} = 0
\label{eq:cont_1}
\end{equation}
\begin{equation}
\rho \, \frac{\partial {\mathbf v}}{\partial t}
+ \rho {\mathbf v} \cdot \bnabla {\mathbf v} = -\bnabla P + 
\rho {\mathbf g} + \bnabla \cdot \taub
\label{eq:mo_1}
\end{equation}
\begin{equation}
\rho \, \frac{\partial h}{\partial t}
+ \rho {\bf v} \bcdot\bnabla h
=k{\bnabla}^2 T + \tau_{ij}\frac{\partial v_i}{\partial x_j}
\label{eq:en_1}
\end{equation}
where $\rho$ is the fluid density, {\bf v} is the velocity vector,
$ {\mathbf g}$ represents the generic body force, $P$ is the pressure,
{$\taub$} is the stress tensor, $h$ is the enthalpy per unit mass, $T$
is the temperature, and $k$ is the thermal conductivity.  
The term containing the stress tensor $\taub$ in equation
(\ref{eq:en_1}) represents the internal heat generated by the viscous 
dissipation (Einstein notation of summation over repeated indeces is
used).  
In this study, for simplicity, the latent heat release due to
crystallization is not considered, the enthalpy is simply given by the
product of a constant specific heat times the temperature. Moreover we
neglect any possible effect due to the buoyancy. 
Under these assumptions and considering a Newtonian relationship
between stress tensor and strain-rate 
($\tau_{ij}=\mu(\partial v_i/\partial x_j + 
\partial v_j/\partial x_i)$), 
equations~(\ref{eq:cont_1}),~(\ref{eq:mo_1}) and (\ref{eq:en_1}) can 
be easily expressed in dimensionless form as:   
\begin{equation}
\displaystyle
\frac{\partial u_i}{\partial \xi_i} = 0
\label{mass}
\end{equation}
\begin{equation}
\displaystyle
\frac{\partial u_i}{\partial \hat{t}} +  
u_j\frac{\partial u_i}{\partial\xi_j} = \frac{1}{Fr}\hat{g}_i
- \frac{\partial p}{\partial \xi_i} 
+ \frac{1}{Re}\frac{\partial}{\partial \xi_j} 
\left[e^{-\Theta} \left(\frac{\partial u_i}{\partial\xi_j} + 
\frac{\partial u_j}{\partial\xi_i}\right)\right]
\label{mom}
\end{equation}
\begin{equation}
\displaystyle
\frac{\partial\Theta}{\partial\hat{t}} + 
 u_j\frac{\partial\Theta}{\partial\xi_j} = 
\frac{1}{Pe} \frac{\partial}{\partial\xi_j} 
\frac{\partial \Theta}{\partial \xi_j} +
\frac{Na}{Pe} \frac{e^{-\Theta}}{2} \left( 
\frac{\partial u_i}{\partial\xi_j} +  
\frac{\partial u_j}{\partial\xi_i}  \right)^2 
\label{ene}
\end{equation}
where $\hat{t}=t U_*/H$ is the dimensionless time,
$(\xi_1,\xi_2)=(x/H,z/H)$ are the longitudinal and transversal
dimensionless coordinates, $(u_1,u_2) = (v_x/U_*,v_z/U_*)$  represent  
the dimensionless field velocities (scaled with the characteristic
velocity $U_*$),  $\Theta=\beta(T-T_0)$ the dimensionless temperature, 
$(\hat{g}_1,\hat{g}_2)=(g_x/|{\mathbf g}|,g_z/|{\mathbf g}|)$ indicate
the dimensionless body force field (from here on-wards we set $g_z=0$)
and $p=P/(\rho U_*^2)$ is the dimensionless pressure (Einstein
convention of summation over repeated indeces is used). The meaning of
the usual characteristic dimensionless numbers is reported  in
Table~\ref{simb}.  \\
Due to the symmetry of the channel and of the boundary conditions, we
investigate only half of the channel ($0\le \xi_2 \le 1$). At the
walls the boundary conditions are given by no-slip velocity and
isothermal temperature: $u_i=\Theta=0$ at $\xi_2=1$ and by
$u_2=\partial u_1/\partial \xi_2 = \partial \Theta/\partial \xi_2=0$
at $\xi_2=0$. At the inlet we assume free flow conditions and the fluid
temperature to be the same as the wall temperature:
$\Theta_{in}=0$. As initial conditions, the velocity and temperature
are set equal to zero. \\    
\begin{table}
\hspace{0.0cm}
\begin{tabular}{lclr|clr}
\hline
 Name & Symbol & Definition & Value &  Symbol & Definition & Value\\
\hline
Reynolds number   & $Re_*$ & $\rho U_*H/\mu_0$    & 4.5  &
$Re$ & $\rho UH/\overline{\mu}$    & 119.4 \\    
Nahme number      & $Na_*$ & $\beta\mu_0 U_*^2/k$ & 14.4 &
 $Na$ & $\beta\overline{\mu}U^2/k$  & 2400 \\
Froude number     & $Fr_*$ & $U_*^2/(g_x H)$         & 1.5  & 
$Fr$ & $U^2/(g_x H)$                  & 412  \\
P\'eclet number     & $Pe_*$ & $\rho c_pU_*H/k$     & 450  &
$Pe$ & $\rho c_p UH/k$             & 7400\\ 
Aspect ratio     & $a_r$ & $H/L$               & 3/100 &
$a_r$ & $H/L$               & 3/100\\
%
%
\hline
\end{tabular}
\caption{Typical dimensionless numbers. The calculated values on the
  left side are based on the mean Poiseuille velocity  
  $U_P = \rho g_x H^2/(3\mu_0)$.
  The calculated values on the right side are instead based the mean
  velocity $U=(\int_0^H v_x dz)/H$ and mean viscosity
  $\overline{\mu}=(\mu_0\int_0^H \exp(-\Theta))/H dz$.}    
\label{simb}
\end{table}
Considering the geometry of \figref{fig:schema} and the isothermal
case without viscous heating effects, the Navier-Stokes equations of a
viscous liquid driven by a body force $g_x$ admit a simple solution
\citep{lanlif94}: 
\begin{equation}
\mu_0\displaystyle\frac{d^2 v_x}{d z^2} +\rho g_x= 0
\qquad   \displaystyle\frac{dP}{dz} = 0
\label{grav}
\end{equation}
In this case, the mean velocity is $U_P = \rho g_x H^2/(3\mu_0)$. \\
From this point on-wards,  we use starred symbols to indicate the
dimensionless number  based the characteristic velocity $U_P$,
i.e. we set $U_*=U_P$, while the un-starred numbers are based on the mean
velocity $U=(\int_0^H v_x dz)/H$, i.e. we set $U_*=U$ (see
Table~\ref{simb}). \\
The parameter values used in the DNS and reported in Table~\ref{simb}
are chosen in order  to perform the computation in a reasonable time,
maintaining the system in the regime with $Re < {\mathcal O}(10^2)$,
$Na  \gg 1$, $Pe \gg 1$, $Pr  \gg 1$ and $H/L\ll 1$.
To fully simulate the flow field evolution when viscous heating
effects are very important, there is a need to solve all the involved
length scales of the problem: from the integral length $H$ up to the
smallest characteristic length-scale. The smallest scales correspond
to a thin layer of the order of $Gz^{-1/2}(\ln Na)^{-1}$ in which the
velocity changes from near zero by the wall to near its core value
($Gz=Pe \times H/L$ indicates the Graetz number) as shown by
\citet{pea77} in the asymptotic limit of very large $Na$ and $Gz$.    
\section{Stability analysis\label{stab}}
The stability of a fully developed steady plane Couette flow was
recently re-examined by \citet{yuewen96}, who improved the results
previously obtained by \citet{sukgol73}. The plane Couette flow shows
two different instability modes: one arising in the non-viscous
limit, and the other due to the viscosity stratification.
As far as the last instability mode is concerned, it was demonstrated
that the critical Reynolds number, above which the flow becomes
turbulent, decreases as the Nahme number increases, that is as the
viscous heating increases \citep{yuewen96}.\\
Viscous heating effects on flow stability have been recently
investigated experimentally by \citet{whimul2000}, who have shown 
that above a critical Nahme number an instability appears at a
Reynolds number one order of magnitude lower than the corresponding
Reynolds number predicted for isothermal flow (in these experiments,
the authors use a temperature-dependent fluid, i.e. glycerin, and a
Taylor-Couette device which allows the tracking of the vortices by a
laser particle tracer). 
\par
When the viscous heating is relevant ($Na \gg 1$) and the thermal
length is much greater than the mechanical one, the temperature
profile, which is characterized by a narrow peak near the channel
wall, is drastically different from the corresponding profile of a
thermally steady fully developed flow
\citep{pea77,ock79,cosmac2003}. Assuming slow longitudinal
variations of velocity and temperature, we now study the linear
stability of a thermally developing flow belonging to the important
regime with $L_t/H=Pe\gg 1$, $L_t/L_m=Pr\gg 1$, $Gz \gg 1$ that
typically characterizes magma flows \citep{wyllis95}. In this regime
it is legitimate to use a lubrication approximation.    
\subsection{Linear stability \label{lin}}
For the investigation of the linear stability we use the method of
small perturbations (normal-mode analysis). The base velocity,
temperature, viscosity and pressure fields are perturbed by
two-dimensional, infinitesimal disturbances. Each variable
($u_i,\Theta,\mu,p$) is given by a steady part plus a small deviation
from the steady state:    
\begin{equation}
\begin{array}{c}
u_{1}(\xi_1,\xi_2,t)=\overline{u}_1(\xi_2) +
\tilde{u}_1(\xi_1,\xi_2,t)\\   
u_{2}(\xi_1,\xi_2,t)=\tilde{u}_2(\xi_1,\xi_2,t)\\ 
p(\xi_1,\xi_2,t)=\overline{p}(\xi_1)+\tilde{p}(\xi_1,\xi_2,t)\\
\Theta(\xi_1,\xi_2,t)=\overline{\Theta}(\xi_1,\xi_2)+
\tilde{\Theta}(\xi_1,\xi_2,t)\\
\nu(\xi_1,\xi_2,t)=\overline{\nu}(\xi_1,\xi_2) +
\tilde{\nu}(\xi_1,\xi_2,t) 
\label{variabili}
\end{array}
\end{equation}
where the overbar symbol indicates the steady part, the tilde the
perturbation, and $\nu=\mu/\mu_0=e^{-\Theta}$ is the
dimensionless viscosity. 
In the (\ref{variabili}), the steady part of temperature and
viscosity depend on the streamwise coordinate $\xi_1$ while the mean
flow is assumed not to vary appreciably with $\xi_1$ over an 
instability wavelength. This means that we study the thermally
developing flow by making the so called quasi-parallel-flow
approximation ($\overline{u_2}\simeq 0$). I.e. one examines the
stability of a model flow having the same streamwise velocity profile
as the real spatially inhomogeneous flow at the selected spatial
location. Since we treat the stability of those systems in the limit  
$Pe \gg 1$ and $Pr \gg 1$,  with the characteristic length $L_t$ much
greater than the other typical mechanical length scales
\citep{pea77,ock79,cosmac2003}, this assumption is legitimate. 
In this regime it is also legitimate to assume that the base flow
satisfies a system of equations similar to that introduced by
\citet{pea77}. At a fixed distance from the inlet, we consider the
following steady equations:   
\begin{equation}
\begin{array}{l}
 \int_0^1 \overline{u}_1 d\xi_2 =1\\
\displaystyle
\frac{\partial \overline{u}_1}{\partial \xi_2} =
Re\left(\frac{\partial \overline{p}}{\partial\xi_1}- 
\frac{\hat{g}_1}{Fr}\right)\xi_2 e^{\overline{\Theta}} \\
\displaystyle
Pe~\overline{u}_1\frac{\partial \overline{\Theta}}{\partial\xi_1} = 
\frac{\partial^2 \overline{\Theta}}{\partial \xi_2^2} + 
Na\left(\frac{\partial \overline{u}_1}{\partial\xi_2}\right)^2 
e^{-\overline{\Theta}}
\label{eqstaz}
\end{array}
\end{equation}
with geometry and coordinate system showed in
figure~\ref{fig:schema}. As boundary conditions we consider
$u_i=\Theta=0$ at $\xi_2=\pm 1$ whereas at the inlet ($\xi_1 = 0$) we
assume a parabolic velocity profile and an uniform temperature
($\overline{\Theta}=0$).  
Equations (\ref{eqstaz}) were solved by a finite-difference method 
with an implicit scheme for the integration along direction $\xi_1$;
the pressure gradient was iteratively adjusted at each step in order
to satisfy mass conservation. \\
In the following, we study the linear stability of the base velocity and
temperature profiles given by (\ref{eqstaz}). Since the variations
with $\xi_1$ depend upon the coupling with the energy equation through
the viscosity, we consider slow temperature variations with $\xi_1$
\citep{pea77}.  Substituting (\ref{variabili}) into the
equations~(\ref{mass}), (\ref{mom}), (\ref{ene}), subtracting the base
flow solutions of (\ref{eqstaz}) and linearizing, we obtain: 
\begin{equation}
\frac{\partial \tilde{u}_1}{\partial \xi_1} + 
\frac{\partial \tilde{u}_2}{\partial \xi_2} = 0
\label{masstab}
\end{equation}
\begin{equation}
\begin{array}{l}
\displaystyle
\frac{\partial \tilde{u}_1}{\partial \hat{t}} +
\overline{u}_1\frac{\partial \tilde{u}_1}{\partial \xi_1} +
\tilde{u}_2\frac{d \overline{u}_1}{\partial \xi_2} =
- \frac{\partial \overline{p}}{\partial\xi_1} + 
\frac{\overline{\nu}}{Re}\left(
\frac{\partial^2 \tilde{u}_1}{\partial \xi_1^2} 
+  \frac{\partial^2 \tilde{u}_1}{\partial \xi_2^2}\right) +\\
\displaystyle
\frac{1}{Re}\frac{d\overline{\nu}}{d\xi_2}
\left(\frac{\partial \tilde{u}_1}{\partial \xi_2} + 
\frac{\partial \tilde{u}_2}{\partial \xi_1}\right) +
\frac{1}{Re}\frac{d\overline{u}_1}{d\xi_2}
\frac{\partial\tilde{\nu}}{\partial\xi_2} 
+ \frac{\tilde{\nu}}{Re}\frac{d^2\overline{u}_1}{d\xi_2^2}
\label{mom1stab}
\end{array}
\end{equation}
\begin{equation}
\displaystyle
\frac{\partial \tilde{u}_2}{\partial \hat{t}} +
\overline{u}_1\frac{\partial \tilde{u}_2}{\partial \xi_1} =
- \frac{\partial\tilde{p}}{\partial\xi_2} + 
\frac{2}{Re}\frac{d\overline{\nu}}{d\xi_2}
\frac{\partial\tilde{u}_2}{\partial\xi_2}  
+ \frac{\overline{\nu}}{Re}\left(
\frac{\partial^2 \tilde{u}_2}{\partial \xi_2^2} 
+  \frac{\partial^2 \tilde{u}_2}{\partial \xi_1^2}\right) 
+ \frac{1}{Re}\frac{d\tilde{\nu}}{d\xi_1}\frac{d\overline{u}_1}{d\xi_2} 
\label{mom2stab}
\end{equation}
\begin{equation}
\begin{array}{c}
\displaystyle
Pe\left(\frac{\partial\tilde{\Theta}}{\partial\hat{t}} + 
\overline{u}_1\frac{\partial \tilde{\Theta}}{\partial\xi_1} +
\tilde{u}_1\frac{\partial \overline{\Theta}}{\partial\xi_1} +
\tilde{u}_2\frac{\partial \overline{\Theta}}{\partial\xi_2} \right) =
\frac{\partial^2\tilde{\Theta}}{\partial\xi_1^2} +
\frac{\partial^2\tilde{\Theta}}{\partial\xi_2^2} +\\
\displaystyle
2\overline{\nu}Na\frac{\partial\overline{u}_1}{\partial\xi_2}
\left(\frac{\partial\tilde{u}_1}{\partial\xi_2} + 
\frac{\partial\tilde{u}_2}{\partial\xi_1}
+\frac{\tilde{\nu}}{\overline{\nu}}
\frac{\partial\overline{u}_1}{\partial\xi_2}\right)
\label{enestab}
\end{array}
\end{equation}
Equations~(\ref{masstab}),~(\ref{mom1stab}),~(\ref{mom2stab}) and
(\ref{enestab}) are similar to those analyzed by \citet{pinlia95}
who investigated how a variable viscosity affects the stability of the
system. In this study we account for the longitudinal variation of the
base temperature ($\partial \overline{\Theta}/\partial\xi_1$) which
was not considered by \citet{pinlia95} and we also introduce new terms
on the right side of equation~(\ref{enestab}) related to the viscous
heating.\\  
In order to eliminate the continuity equation (\ref{masstab}), we
introduce a perturbation streamfunction $\tilde{\psi}$:
\begin{equation}
\tilde{u}_1=\frac{\partial \tilde{\psi}}{\partial\xi_2}
\qquad\tilde{u}_2=-\frac{\partial \tilde{\psi}}{\partial\xi_1}
\label{stream}
\end{equation}
Moreover we assume that all perturbations have temporal and
spatial dependence of the form:
\begin{equation}
(\tilde{\psi},\tilde{p},\tilde{\Theta},\tilde{\nu}) = 
[\phi(\xi_2),f(\xi_2),\theta(\xi_2),\Lambda(\xi_2)]
\ e^{i\alpha(\xi_1-c\hat{t})} 
\label{waves}
\end{equation}
where $\alpha$ is the wavenumber, $c$ is the complex perturbation
velocity and $\phi,f,\theta,\Lambda$ indicate the disturbance
amplitudes.  \\
Substituting equations~(\ref{stream}) and (\ref{waves}) into the 
(\ref{masstab}), (\ref{mom1stab}), (\ref{mom2stab}) and
(\ref{enestab}), and eliminating the pressure disturbance term by
cross differentiation and subtraction, we obtain the final stability
equations: 
\begin{equation}
\begin{array}{c}
i\alpha Re\left[(\overline{u}-c)(\phi{''}-\alpha^2\phi)-
\overline{u}^{''}\phi\right] =
\overline{\nu}(\phi^{iv}-2\alpha^2\phi^{''}+ \alpha^4\phi)+\\
2\overline{\nu}'(\phi{'''}-\alpha^2\phi')+
\overline{\nu}{''}(\phi{''}+\alpha^2\phi)+
\overline{u}'(\Lambda{''}+\alpha^2\Lambda)+2\overline{u}{''}\Lambda' 
+\overline{u}{'''}\Lambda
\label{orrsomm}
\end{array}
\end{equation}
\begin{equation}
i\alpha Pe\left[(\overline{u}-c)\theta-\phi\overline{\Theta}'+
\phi'\frac{\partial \overline{\Theta}}{\partial \xi_1}\right] =
(\theta{''}-\alpha^2\theta)+
2\overline{\nu} Na\left[(\phi{''}+ \alpha^2\phi)+ 
\frac{\Lambda}{\overline{\nu}}\overline{u}'\right]\overline{u}'  
\label{stabtemp}
\end{equation}
where  for simplicity with $\overline{u}$ we indicate the velocity
base flow $\overline{u}_1$ and the symbol prime indicates
differentiation with respect to $\xi_2$.  
Viscosity perturbation $\Lambda$ can be expressed in terms of
temperature fluctuations by the Taylor expansion of
(\ref{eq:viscosity_exponential}), and neglecting nonlinear terms:
\begin{equation}
\Lambda=-\theta\bar{\nu}
\label{viscpert}
\end{equation}
obtaining the two final governing stability equations for $\phi$ and 
$\theta$. Finally, as boundary conditions for (\ref{orrsomm}) and
(\ref{stabtemp}), we consider: 
\begin{equation}
\phi=0, \quad \phi'=0, \quad \theta=0 \qquad \mbox{at} \quad 
\xi_2=\pm 1
\label{bc}
\end{equation}
We note that equation~(\ref{orrsomm}) reduces to the classical
Orr-Sommerfeld equation when $\bar{\nu}=1$ and the
equation~(\ref{stabtemp}) reduces to that used by \citet{pinlia95}
when both $Na=0$ and $\partial \overline{\Theta}/\partial \xi_1=0$.
\subsection{Solution method and stability results}
Classical flow stability problems are usually approached in two ways:
temporal and spatial. In the former case, it is assumed that small
disturbances evolve in time from some initial spatial distribution. In
this case, for an arbitrary positive real value of $\alpha$, the
complex eigenvalue $c=c_R+i c_I$ and the corresponding eigenfunctions
$\phi$ and $\theta$ are obtained. If $c_I=\Imag (c)$ is negative then
the flow is temporally stable, otherwise it is unstable.  \\ 
The spatial analysis is focused on the spatial evolution of a time
periodic perturbation at a fixed position in the flow. This study
requires the solution of a nonlinear eigenvalue problem in $\alpha$,
which is assumed complex $\alpha=\alpha_R+i\alpha_I$ with a prescribed
real $c=c_R$. The disturbances grow for $\Imag (\alpha)< 0$ and decay
for  $\Imag (\alpha)> 0$.  \\
The choice between spatial and temporal study depends on the nature of
the flow instability considered \citep[see ][for a general
  review\ ]{huemon90}. Moreover quasi-parallel flows may contain
different region with different stability characteristics.\\
In the present paper, a temporal stability analysis of the profiles at
a selected set of distances from the inlet, has been performed. 
This analysis is adequate for studying the so-called absolute
instabilities (i.e. when the perturbation contaminates the entire flow
both upstream and downstream of the source location).
\subsubsection{Temporal stability study}
\begin{figure}
\hspace{-0.7cm}
\includegraphics[angle=-90,width=0.5\hsize]{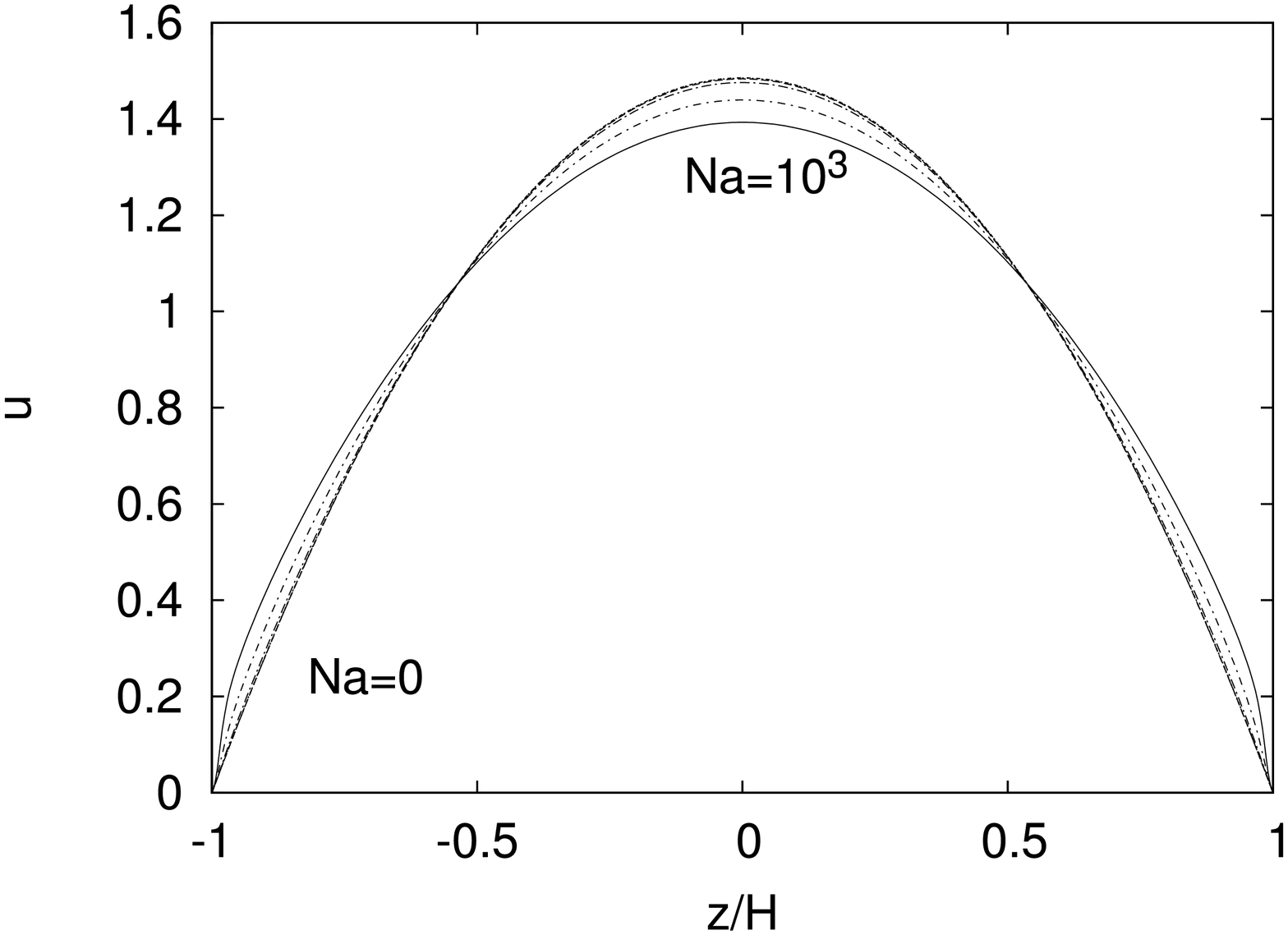}
\includegraphics[angle=-90,width=0.5\hsize]{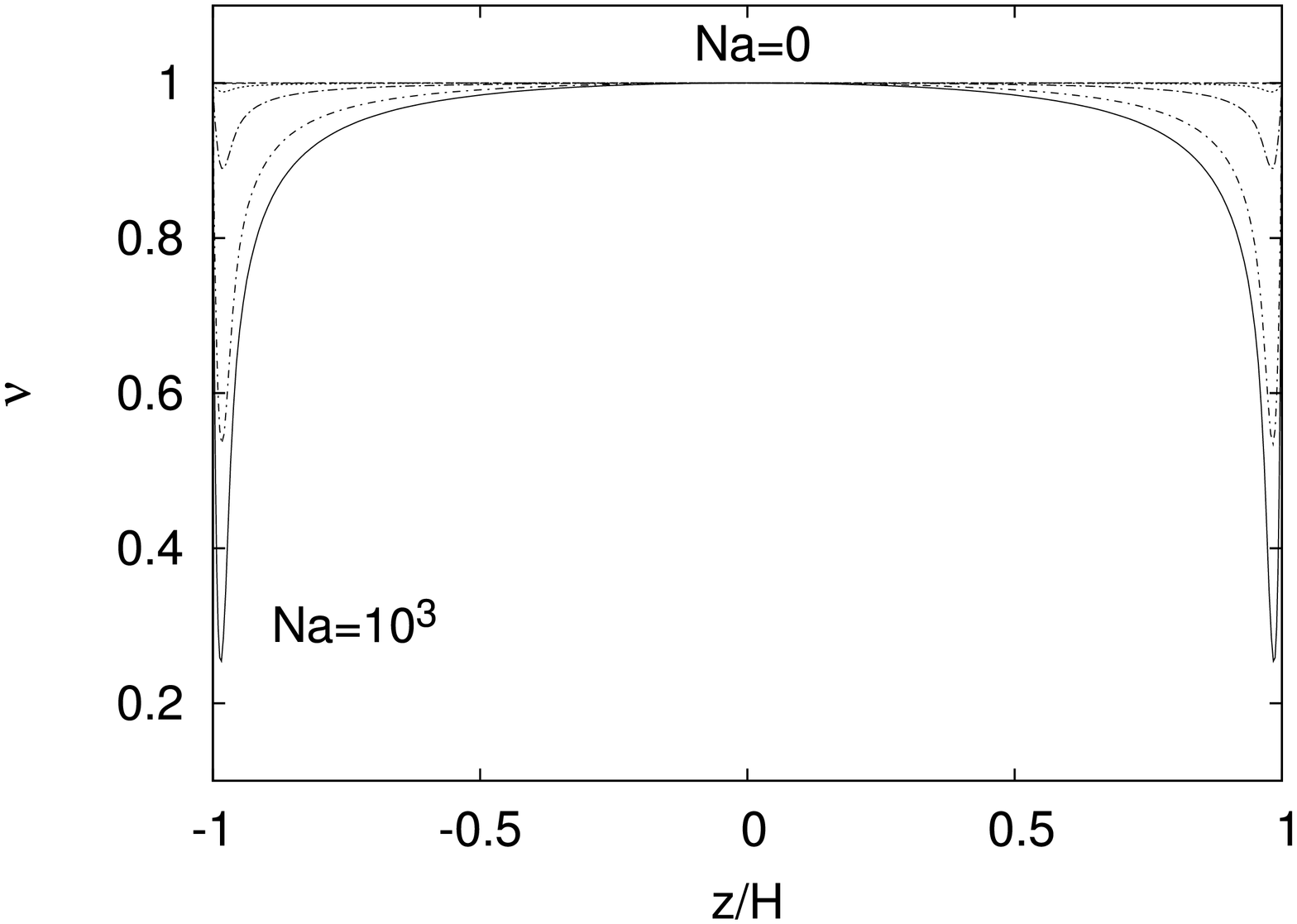}
\caption{Base velocity profiles (on the left) and base viscosity
  profiles (on the right) at $\xi_1^*=100$ for  $Pe=10^7$ and for
  $Na=0; 1; 10; 100; 1000$. Here velocity profiles are normalized with 
  respect to the mean velocity $U$.}    
\label{baseflow}
\end{figure}

The problem formulated in the \S~\ref{lin} is solved using a
Chebyshev collocation technique, expanding the functions $\phi$ and
$\theta$ in series of Chebyshev polynomials of order N. The 2(N+1)
coefficients are considered as unknowns and they are evaluated by the   
collocation technique applied at points 
$\xi_{2,i}=\cos(\frac{\pi i}{N-3}) $ with $i=0,1,2,3... N-3$ and
imposing the six boundary conditions (\ref{bc}) at $\xi_2=\pm 1$.
This method allows us to define a system of $2(N+1)$ equations in
$2(N+1)$ unknowns which can be written as a generalized eigenvalue
problem of the type Ax=cBx. The final system was solved using the
LAPACK routine ZGGEV. Typically, setting $N=70$ and $N=80$ permits a
satisfactory convergence in the computation of the eigenvalues.     
In order to test the above described computational implementation we
compared the obtained eigenvalues in the limit $\Theta\rightarrow 0$,
with \citet{ors71}'s results (considering Orszag's definitions, our
$c$ is 1.5 times Orszag's $c$ while Orszag's $Re$ is 1.5 times our
$Re$). Table \ref{os} shows that eigenvalues we calculated for
isothermal limits are very close to those obtained by \citet{ors71}.\\   
\begin{table}
\hspace{0.0cm}
\begin{tabular}{lll}
\hline
 Mode Number & Eigenvalues by \citet{ors71} & Our eigenvalues for
 $\Theta = 0$ \\ 
\hline
1   & 0.23752649 + 0.00373967 i & 0.237526311 + 0.00373795 i\\
2   & 0.96463092 - 0.03516728 i & 0.964629174 - 0.03516535 i \\
3   & 0.96464251 - 0.03518658 i & 0.964643595 - 0.03518749 i\\
4   & 0.27720434 - 0.05089873 i & 0.277207006 - 0.05089868 i\\
5   & 0.93631654 - 0.06320150 i & 0.936328259 - 0.06320707 i\\
... & ...                      & ...\\
\hline
\end{tabular}
\caption{Least stable eigenvalues $c=c_R+i c_I$ calculated in this
  work in the isothermal limit  compared with the \citet{ors71}'s
  results for $Re= 10^4$ and $\alpha =1$. $N=70$ was set.}      
\label{os}
\end{table}
As far as the base flow is concerned, we considered a fixed distance
from the inlet $\xi_1^*$ and a given P\'eclet number $Pe$. As shown in
figure~\ref{baseflow} for $Pe=10^7$, as the Nahme number increases,
velocity distributions deviate from parabolic profile and
dimensionless viscosity drops near the walls. 
\par
\begin{figure}
\hspace{-0.7cm}
\includegraphics[angle=-90,width=0.5\hsize]{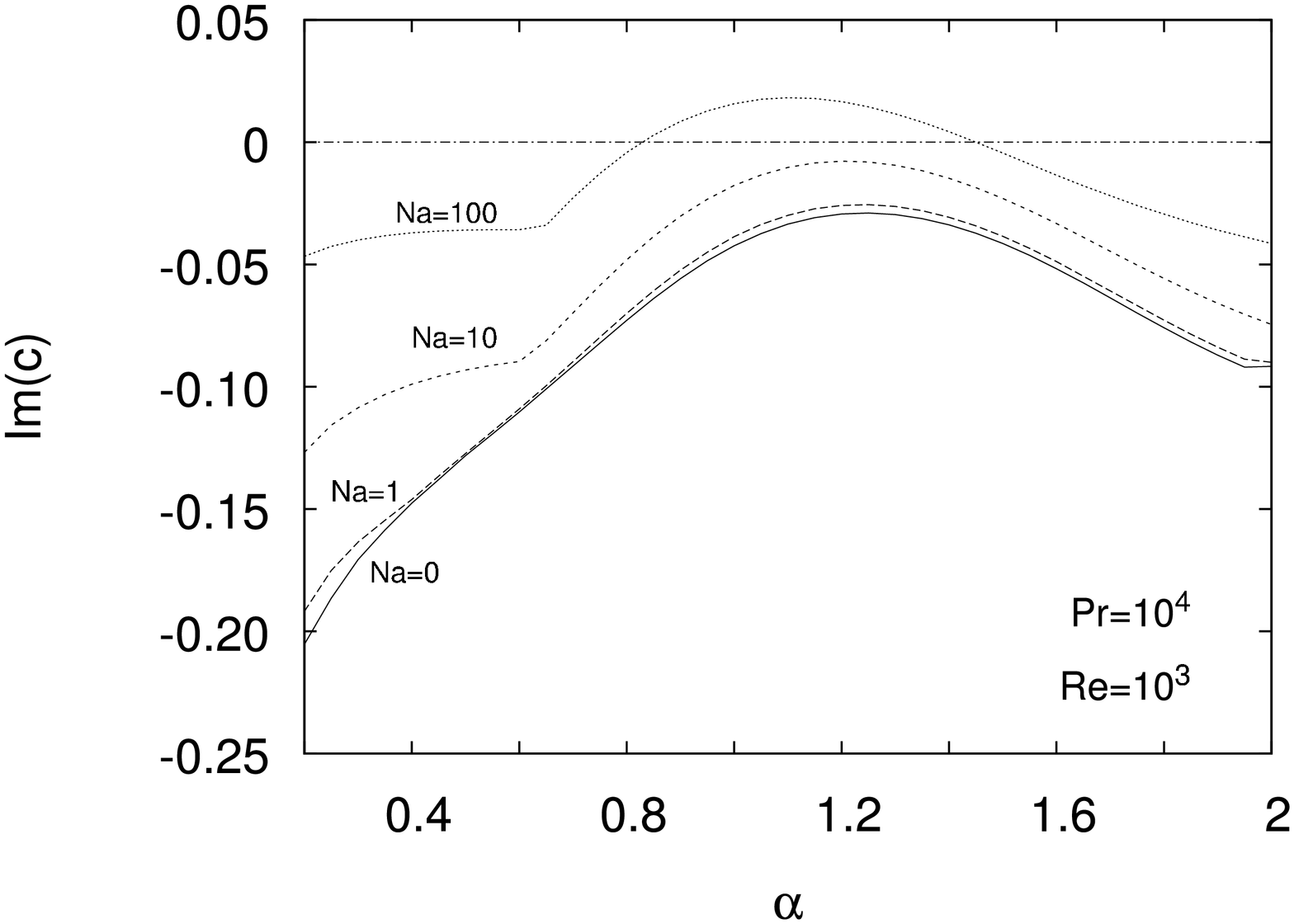}
\includegraphics[angle=-90,width=0.5\hsize]{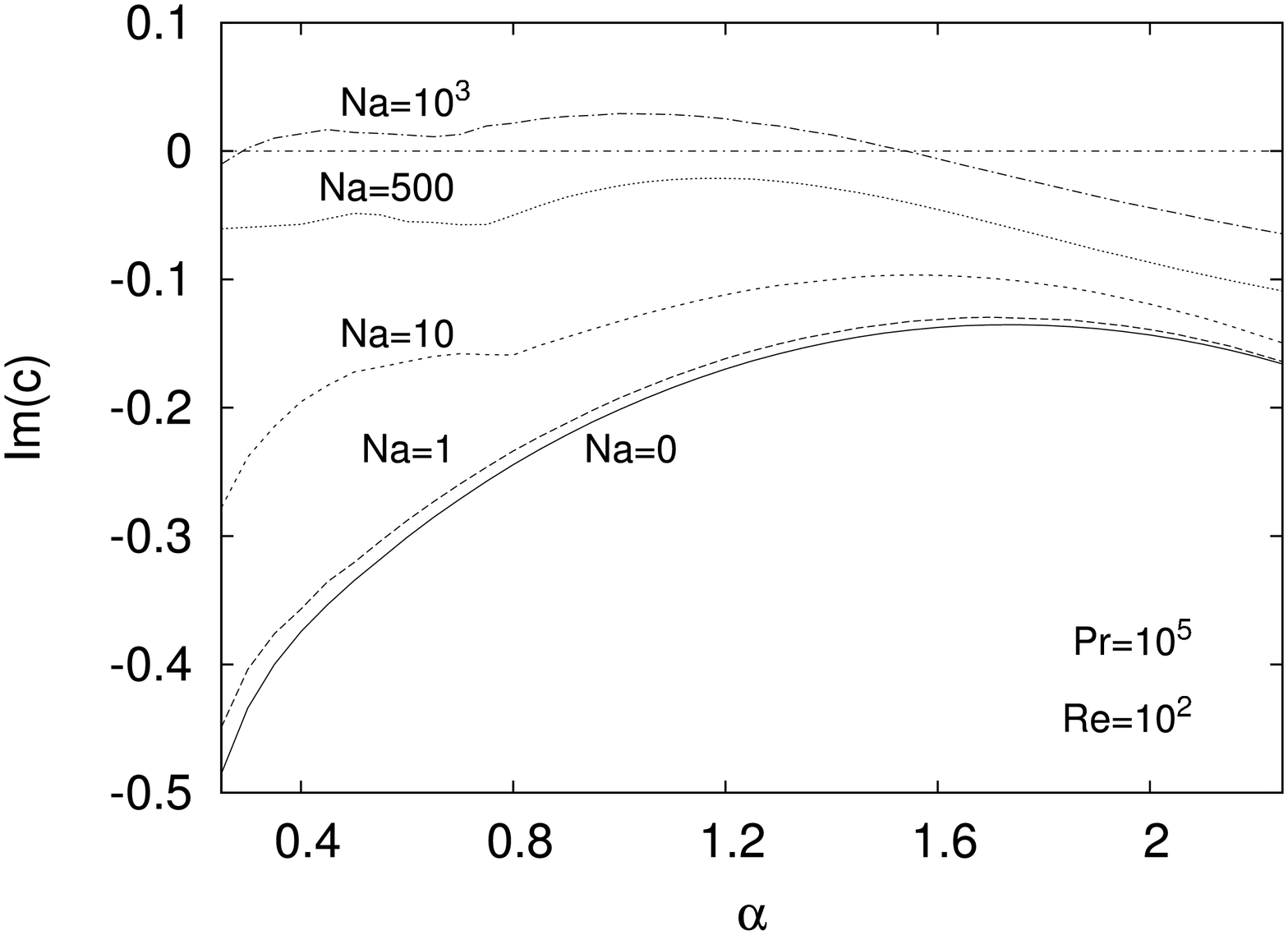}
\caption{Imaginary part of the complex perturbation velocity vs $\alpha$: 
  $Pr=10^4$, $Re=10^3$ (on the left) and $Pr=10^5$ and $Re=10^2$ (on
  the right). For both cases a distance from the inlet of
  $\xi_1^*=100$ was set.}  
\label{stabci}
\end{figure}

The stability analysis shows that viscous heating in fluids with
temperature-dependent viscosity is destabilizing. In fact in the cases
studied, for a given $Pr$ there is a critical Nahme number $Na_c$
above which the flow is unstable at any $Re$, i.e. the critical
Reynolds number $Re_c$ decreases as the Nahme number $Na$ increases.
Two clear examples of this are shown in figure~\ref{stabci} where, for
different values of $Na$, the imaginary part of the eigenvalue $c$ is
plotted as a function of the wavenumber $\alpha$ at a distance
$\xi_1^*=100$ from the inlet and for $(Re=10^2, Pr=10^5)$ and
$(Re=10^3, Pr=10^4)$, respectively. From these plots, it is evident
that increasing the Nahme number, the imaginary part of the complex
perturbation velocity tends to increase until becomes positive. For
instance, for $Pr=10^5$ and $Re=10^2$ the flow becomes unstable for
$Na\lesssim 10^3$ while at $Pr=10^5$ and $Re=10^3$ the flow is
unstable for $Na\gtrsim 10$. The same behaviour was observed with
lower Reynolds number where the flow becomes unstable at larger $Na$.     
\par
Beside the investigation of the role of the Nahme and Reynolds numbers
on the flow stability, a more deepened parametric study of the effects
of the other controlling parameters of the problem, such as
the P\'eclet number and the distance from the inlet, should be
performed. In any case, even considering our preliminary results, it
is clear that viscous heating effects in fluids with
temperature-dependent viscosity are important for the determination of
flow instablities and, without their inclusion, the critical Reynolds
number is generally overestimated.  
\section{Numerical simulation}
\label{numsim}
In this section, we first describe the numerical scheme, then the 
numerical parameters used to solve the equations described above and,
finally, we present the results obtained and discuss them.      
\subsection{Numerical scheme\label{num_scheme}}
To solve equations~(\ref{eq:cont_1}),~(\ref{eq:mo_1}) and
(\ref{eq:en_1}), a fortran code based on the Finite Element Method
(FEM) with the Streamline-Upwind/Petrov-Galerkin (SU/PG) scheme
\citep{brohug82} was used. The enthalpy equation is added in a similar
way as suggested by \citet{heiyu88}.
The solution method is explicit in the velocity and temperature and
implicit in the pressure, which is computed solving a Poisson
equation.\\ 
The domain of interest $\Omega$ is partitioned into a number of non
intersecting elements $\Omega^e$ with $e$=1,2,\dots,$n_{e}$, where
$n_{e}$ is the total number of elements.   
In contrast with the usual Galerkin method, which considers the
weighting functions continuous across the element boundaries, the
SU/PG formulation requires discontinuous weighting functions of the
form: $\tilde w = w + \tilde s $,  where $w$ is a continuous weighting
function (the Galerkin part) and $\tilde s$ is the discontinuous
streamline upwind part. Both $w$ and $\tilde s$ are smooth inside the
element. The upwinding functions $\tilde s$ depend on the
local element Reynolds number (momentum equations) and the local
element P\'eclet number (enthalpy  equation).  The SU/PG weighting
residual formulation of the initial-boundary value problem defined
by equations~(\ref{eq:cont_1}) and (\ref{eq:mo_1}) can be respectively
recasted as:   
\begin{equation}
\sum_e \int_{\Omega_e} \tilde s^p_k \,\frac{\partial v_i}{\partial x_i}
\, d \Omega = 0 
\label{eq:cont_2}
\end{equation}
where $\tilde s^p_k$ is a weighting function which is chosen to be
constant within each element, and discontinuous across the element
boundaries, and 
\begin{equation}
\begin{array}{l}
\displaystyle
\lefteqn{\int_{\Omega} w_k \left( \rho \, \frac{\partial v_i}
  {\partial t} + \rho v_j \, \frac{\partial v_i}{\partial x_j} -
  \rho g_i \right) d \Omega + \int_{\Omega} \sigma_{ij} \,
  \frac{\partial w_k}{\partial x_j} \, d \Omega \, + } \\
\displaystyle
+ \sum_e \int_{\Omega} \tilde s_k^u
\left( \rho\, \frac{\partial v_i}{\partial t}+
\rho v_j \, \frac{\partial v_i}{\partial x_j}
-\frac{\partial \sigma_{ij}}{\partial x_j} -\rho g_i \right) d \Omega =
\int_{\Gamma_{\sigma}} \sigma_{0i} \,w_k \, d\Gamma \label{eq:mom_2}
\end{array}
\end{equation}
where $\tilde s_k^u$ is the upwinding function for the momentum
equation, $\sigma_{ij} = -P\delta_{ij} + \tau_{ij}$ ($\delta_{ij}$ is
the Kronecker symbol). With $\Gamma_{\phi}$ we generally indicate the
boundary surface where the variable $\phi$ is prescribed. \\
In the same way, the weak form of the enthalpy equation
(\ref{eq:en_1}) may be written as: 
\begin{equation}
\begin{array}{l}
\displaystyle
\lefteqn{\int_{\Omega} w_k \left( \rho \,\frac{\partial h}{\partial t}  
+ \rho v_i\,\frac{\partial h}{\partial x_i} \right) d \Omega + 
\int_{\Omega} \left( k\,\frac{\partial T}{\partial x_i}\right)
\frac{\partial w_k}{\partial x_i}\,d\Omega \,+ }  \\
\displaystyle
+\sum_e \int_{\Omega} \tilde s_k^h\left[\rho\,\frac{\partial h}
  {\partial t} + \rho v_i\,\frac{\partial h}{\partial x_i}
-\frac{\partial}{\partial x_i}\!\left(k\,\frac{\partial T} 
{\partial x_i} \right) + \tau_{ij}\,\frac{\partial v_i}{\partial
  x_j} \right] d\Omega = \int_{\Gamma_h} q_0\,d\Gamma
\label{eq:en_2} 
\end{array}
\end{equation}
where, $\tilde s_k^h$ is the upwinding function of the enthalpy
equation and $q_0$ is the heat flux through the boundary surface
$\Gamma_h$. 
In the present work, the velocity field $v$ and the enthalpy
fields are linearly interpolated with multi-linear
iso-parametric interpolation functions using rectangular elements.
The pressure field $P$, instead, is assumed to be constant within each  
element and discontinuous across the element boundaries. \\
Equations \paref{eq:mom_2} and \paref{eq:en_2} yield two algebraic
equations which may be combined in the following one:
\begin{equation}
\mathbf{Ma + Cv + N(v) - GP = F}
\label{eq:mat_1}
\end{equation}
whilst the continuity equation \paref{eq:cont_2} yields to:
\begin{equation}
\mathbf{G^T v = D}
\label{eq:mat_2}
\end{equation}
In the above equations, the vector $\mathbf v$ represents the nodal
values of the velocity $v_i$ and the temperature $T$, whereas the
vector $\mathbf a$ represents the nodal values of the time derivatives 
of the velocity $\dot v_i$ and of the temperature $\dot T$, $\mathbf P$
indicates the pressure field, ${\bf M}$ is the consistent generalised 
mass matrix, ${\bf C}$ and ${\bf N(v)}$  account for the diffusive and 
the nonlinear convective terms respectively, ${\bf F}$ is a
generalised force vector, ${\bf D}$ accounts for the prescribed
velocity at the boundaries, ${\mathbf G}$ is the gradient operator,
and ${\mathbf{G^T}}$ its transpose.
Equations \paref{eq:mat_1} and \paref{eq:mat_2} are integrated in time
starting from the velocity and pressure fields at $t=0$.\\
Convergence of the algorithm is assured when the element Courant   
number $Cr$ satisfies particular conditions depending on the element
Reynolds and P\'eclet numbers. Typically, to assure the convergence,
the element Courant number should satisfy the most restrictive among
the following relations \citep{brohug82}: 
\begin{eqnarray*}
Cr \le 0.8, \qquad & \hbox{if} & \quad
\gamma=\left\{^{\displaystyle Re_{el}}_{\displaystyle Pe_{el}} \ge
100,\right. \\ 
Cr \le \hbox{min}(1,\gamma), \qquad & \hbox{if} & \quad
\gamma=\left\{^{\displaystyle Re_{el}}_{\displaystyle Pe_{el}}  <
100\right.   
\end{eqnarray*}
where $Cr = v \Delta t /\Delta x_i$, with $v$, $\Delta t$ and 
$\Delta x_i$  element velocity, computational time step and
computational grid size respectively, and $\gamma=Re_{el} \mbox{ or }
Pe_{el}$ represents both the element Reynolds and P\'eclet numbers. 
Unfortunately, the above convergence criteria can be very restrictive,
forcing the choice of a very small time step to guarantee  the
convergence of the algorithm.   
\subsection{Numerical parameters\label{num_param}}
\begin{figure}
\includegraphics[angle=0,width=\hsize]{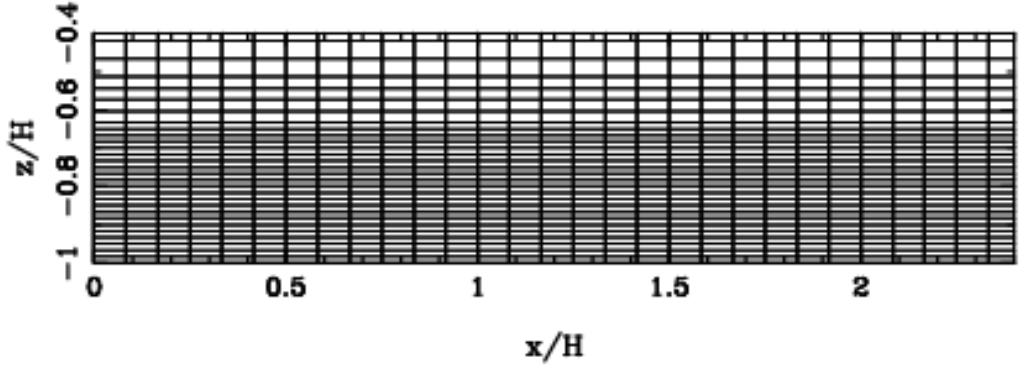}
\caption{Zoom of the bottom-left corner of the computational grid used
  for the simulations. The entire computational domain was discretized
  with $401 \times 38$ rectangular elements. The longitudinal 
  grid size was chosen uniform $\Delta x/H= 8.3\cdot 10^{-2}$ while
  three different sizes were used to form the transverse grid:
  $n_1=23$ elements of size $\Delta z_1/H=1.67\cdot 10^{-2}$  near the
  wall, $n_2=5$ elements of size $\Delta z_2/H=3\cdot 10^{-2}$ in the
  intermediate region, and $n_3=10$ elements of size $\Delta
  z_2/H=4.67\cdot 10^{-2}$ in the central part of the channel.}
\label{grid}
\end{figure}

Since viscous friction is greater near the walls (higher gradients),
it is convenient to use a computational grid finer near the boundaries
and coarser towards the centre of the channel.\\
The computational grid was formed by an uniform horizontal grid size  
$\Delta x/H=8.3\cdot 10^{-2}$ while the vertical mesh size $\Delta\zeta
=\Delta z/H$ is not uniform and consists of three different grid
sizes $\Delta\zeta_1$, $\Delta\zeta_2$, and $\Delta\zeta_3$. The
finest grid size $\Delta\zeta_1=1.67\cdot 10^{-2}$ was set near the
wall where the fields change rapidly and the viscosity is lower,
$\Delta\zeta_2=3\cdot 10^{-2}$ was set in the intermediate region
and finally, $\Delta\zeta_3=4.67\cdot 10^{-2}$  in the central part of
the channel. The computational grid used for the simulations is shown
in figure~\ref{grid}. \\      
A time step of $\Delta \hat{t} = 5\cdot 10^{-4}$  was chosen to
perform the simulation, with one predictor and one corrector
iterations per time step, whereas spatial integration was performed
using a 2x2 Gauss quadrature. All runs were performed in
double-precision arithmetics on a HP-J5600 workstation. \\  
From a practical point of view, an estimation of the minimum grid size
required to well resolve the rapidly varying fields was assumed to be
equal to the smallest scale involved in the problem \citep{pea77}: 
$\Delta\zeta_1^{(0)}= Gz^{-1/2}(\ln Na)^{-1}$  (using some
$Na$ and $Gz$ initial estimations), it was then further reduced to
guarantee the numerical convergence and in such a way that the
numerical  solution does not appreciably depend on the computational
grid size. The final dimensionless numbers used in the DNS are
reported in Table~\ref{simb}.   
\subsection{Results and discussion}
In this section we describe results obtained by the direct numerical
simulation (using the FEM code described in \S~\ref{num_scheme}) for
the values of the parameters reported in \S~\ref{num_param} and
Table~\ref{simb} for a channel flow of length 100/3 $H$-unit.   
\par
In this study we relied on the small numerical round-off errors
present in any numerical simulation to trigger natural modes, but
further cases with e.g. selected harmonic or random input disturbance
should be investigated.  
\par
\begin{figure}
\hspace{-0.12\hsize}
\includegraphics[angle=-90,width=0.6\hsize]{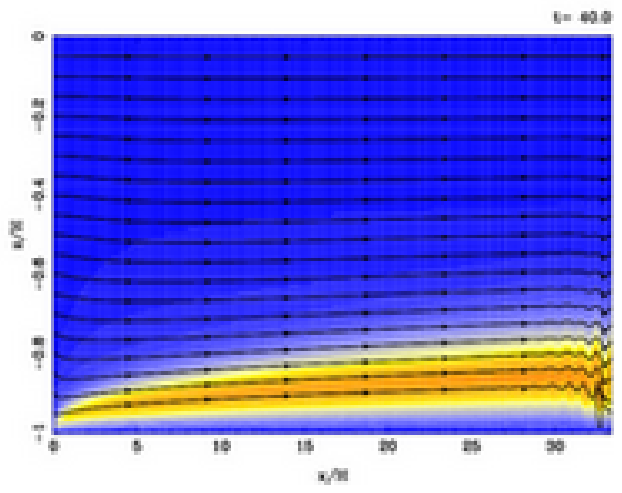}
\hspace{-0.1\hsize}
\includegraphics[angle=-90,width=0.6\hsize]{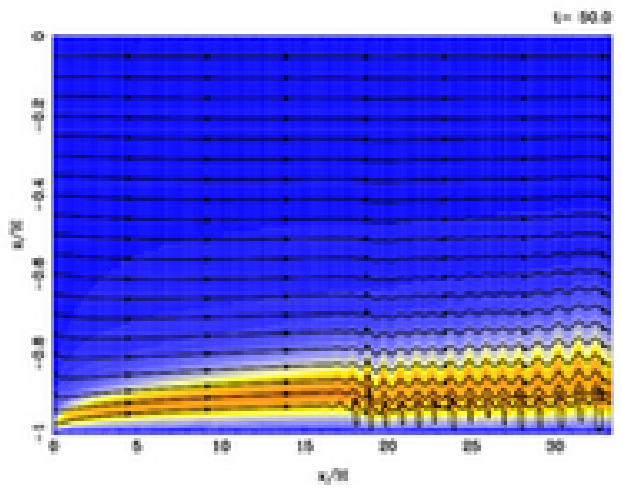}

\hspace{-0.12\hsize}
\includegraphics[angle=-90,width=0.6\hsize]{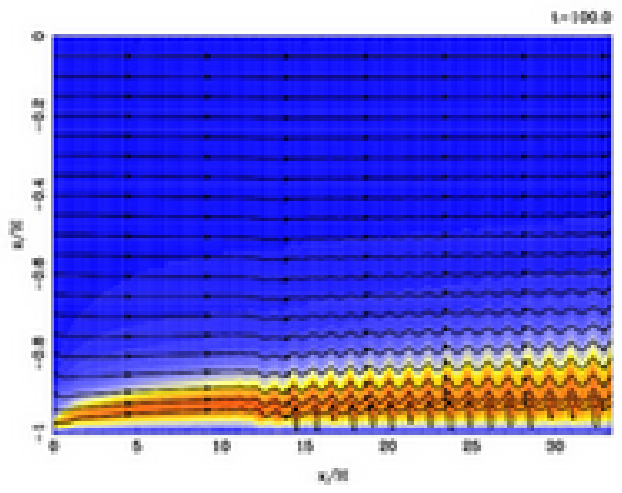}
\hspace{-0.1\hsize}
\includegraphics[angle=-90,width=0.6\hsize]{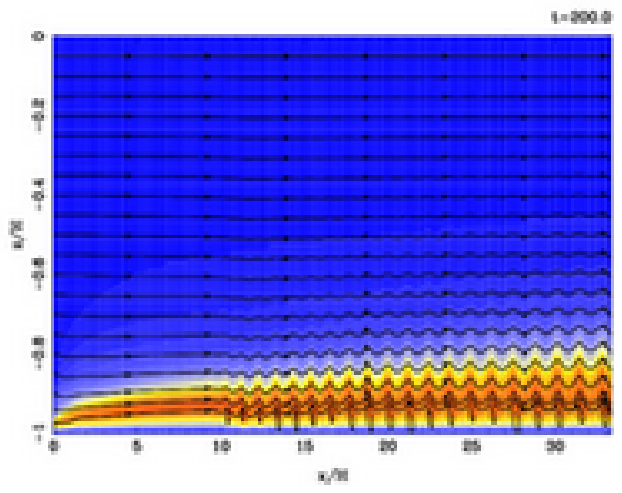}
\caption{Evolution of the dimensionless flow fields. The figures show
  the simulated streamlines with the temperature field as background
  at different times $\hat{t}$. The values reported along vertical
  and horizontal axes indicate the dimensionless distances from the
  channel wall and from the inlet, respectively.
  The blue colour indicates the lowest dimensionless temperature
  ($\Theta=0$) and the dark orange  corresponds to the highest
  temperature ($\Theta=6$). The symbol $t$ on the upper right corner
  indicates the dimensionless time $\hat{t}$. For visualization
  reasons the horizontal axis is contracted with respect to the
  vertical.}     
\label{temp}
\end{figure}
\begin{figure}
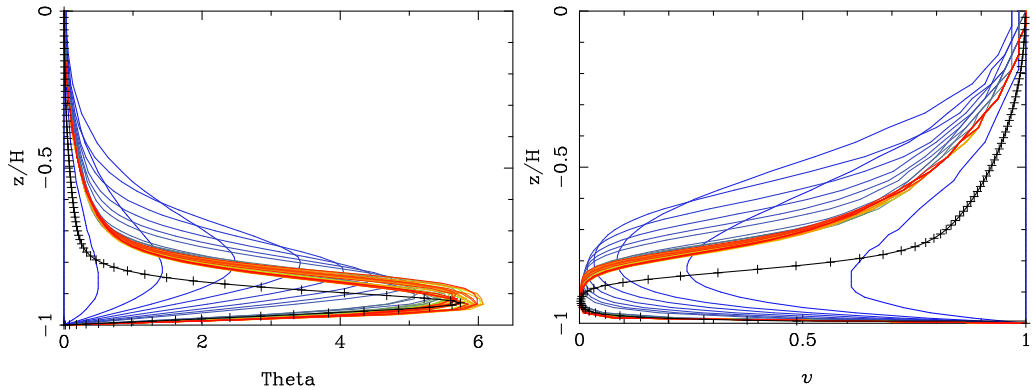

\hspace{-0.6cm}
\includegraphics[angle=-90,width=0.5\hsize]{tempad.eps}
\includegraphics[angle=-90,width=0.5\hsize]{viscad.eps}
\caption{Temporal evolution from $\hat{t}=0$ (blue colour) to
  $\hat{t}=250$ (red colour) of dimensionless temperature profile
  $\Theta$ (on the left) and  dimensionless  viscosity profile $\nu$
  (on the right), at a  dimensionless distance from inlet of about 2/3
  of the tube length ($\xi_1^*=22$). The coding colour maps the time
  evolution from the blue up to red. For comparison, temperature and
  viscosity profiles computed using the lubrication approximation
  (equations~(\ref{eqstaz})) are also reported (full lines with
  crosses).}      
\label{tempad}
\end{figure}
\begin{figure}
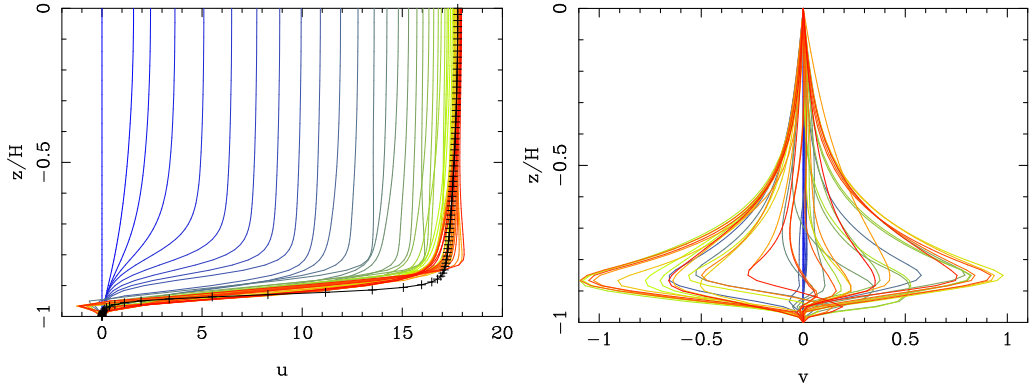

\hspace{-0.6cm}
\includegraphics[angle=-90,width=0.5\hsize]{vxad.eps}
\includegraphics[angle=-90,width=0.5\hsize]{vyad.eps}
\caption{Temporal evolution from $\hat{t}=0$ (blue colour) to
  $\hat{t}=250$ (red colour) of the profile of the dimensionless
  longitudinal velocity $u=u_1=v_x/U_P$ (on the left) and of the
  dimensionless transversal velocity profile  $v=u_2=v_z/U_P$ (on the 
  right), at a dimensionless distance from inlet of about 2/3 of the
  tube lenght ($\xi_1^*=22$). The coding colour maps the time
  evolution from the blue up to red. For comparison, the velocity
  profile computed using the lubrication approximation
  (equations~(\ref{eqstaz})) is also reported (full line with
  crosses).}    
\label{vvad}
\end{figure}

From these simulations we can see that as the time increases the
temperature starts to rise in the region near the outlet because of
viscous heating. At $\hat{t}=\hat{t}_c\approx 40$  an instability is
triggered in this region where the dimensionless temperature $\Theta$
locally becomes greater than $\approx 5$ (see figure~\ref{temp}). For
$\hat{t}>\hat{t}_c$, as viscous heating effects become more
important even in the more internal region, secondary flows appear to
organize themselves into ``coherent structures'' as rotational
flows. This kind of secondary flow looks like roller vortices which
seem to move from the region near the outlet towards the inlet (see
figure~\ref{temp}). Actually this happens because the viscous heating
becomes relevant even in the internal region and the entire flow
becomes unstable. \\  
Figure~\ref{tempad} show the temporal profile evolution at a given
distance (for example at $\xi_1=22$ that is about 2/3 of the channel
length). We can see that $\Theta$, starting with a flat distribution,
gradually increases near the wall forming a profile with a maximum at
a short distance from the boundary. As time increases, this peak
becomes more pronounced ($\Theta_{max}\lesssim 6$) filling, at the
steady state, a narrow shell of values at a shorter distance from the
wall (see figure~\ref{tempad}). As a consequence the dimensionless
viscosity profile $e^{-\Theta}$, strongly decreases in correspondence
of the temperature peak, reaching values much samller than its
initial ones (see figure~\ref{tempad}). The layer where the
viscosity is very small is immediately close to the colder layer
adjacent to the wall and it corresponds to the region where the
vortices appear (see figures~\ref{temp}). \\ 
The longitudinal velocity profile (scaled by $U_P$), starting with a
parabolic distribution, evolves toward a plug profile filling, at the
steady state, a narrow region of values with a plug velocity $\lesssim
18$ (see figure~\ref{vvad}). Figure~\ref{vvad} also shows the
evolution of the transversal dimensionless velocity profiles which,
because of the vortical motions,  near the wall, tend to fill an onion
shape region with the largest fluctuations in corrispondence of the
peak in the temperature profile. For comparison, the profiles computed
using the lubrication approximation (\ref{eqstaz}) are reported in
figures~\ref{tempad} and \ref{vvad}.\\        
\begin{figure}
\includegraphics[angle=0,width=\hsize]{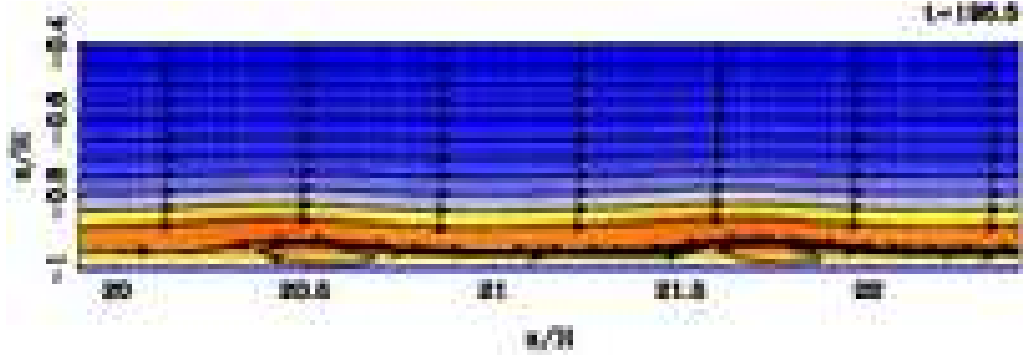}
\caption{Visualization of the flow structures near the channel wall for
  $\hat{t}=195$ at a dimensionless distance from inlet of about 2/3 of
  the channel length. The blue colour indicates the lowest dimensionless 
  temperature ($\Theta=0$) and the darkest orange corresponds to the
  highest temperature ($\Theta=6$).}    
\label{zoom}
\end{figure}
\par
\begin{figure}
\hspace{-0.7cm}
\includegraphics[angle=-90,width=\hsize]{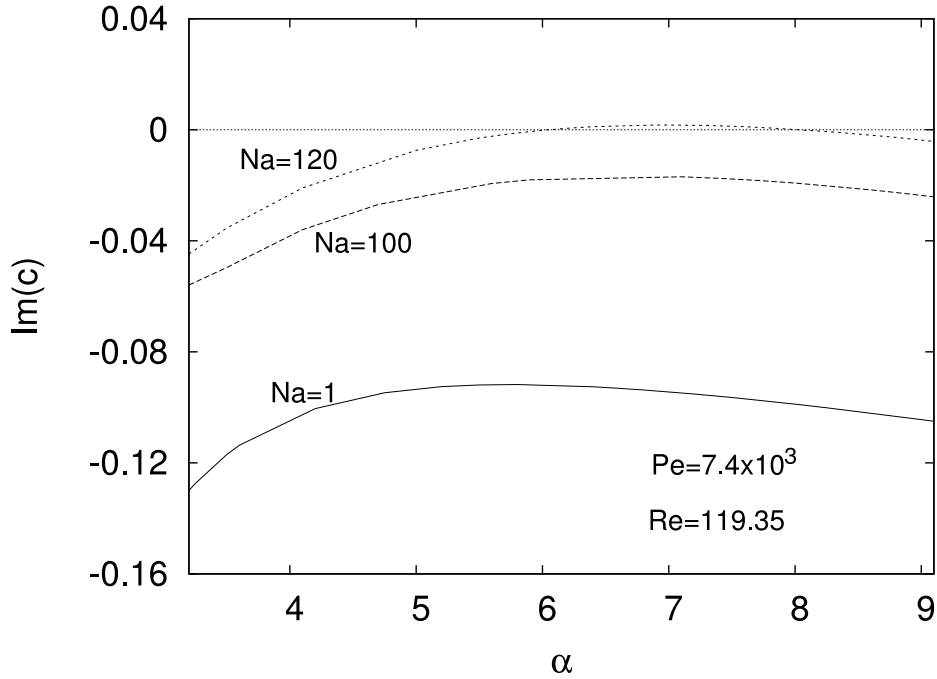}
\caption{Imaginary part of the complex perturbation velocity vs $\alpha$: 
  $Pe=7400$, $Re=119.35$ and $Pr=62$ at a dimensionless distance from
  inlet of about 2/3 of the channel length ($\xi_1^*=21$).}
\label{stabmode}
\end{figure}
\begin{figure}
\hspace{-0.6cm}
\includegraphics[angle=-90,width=\hsize]{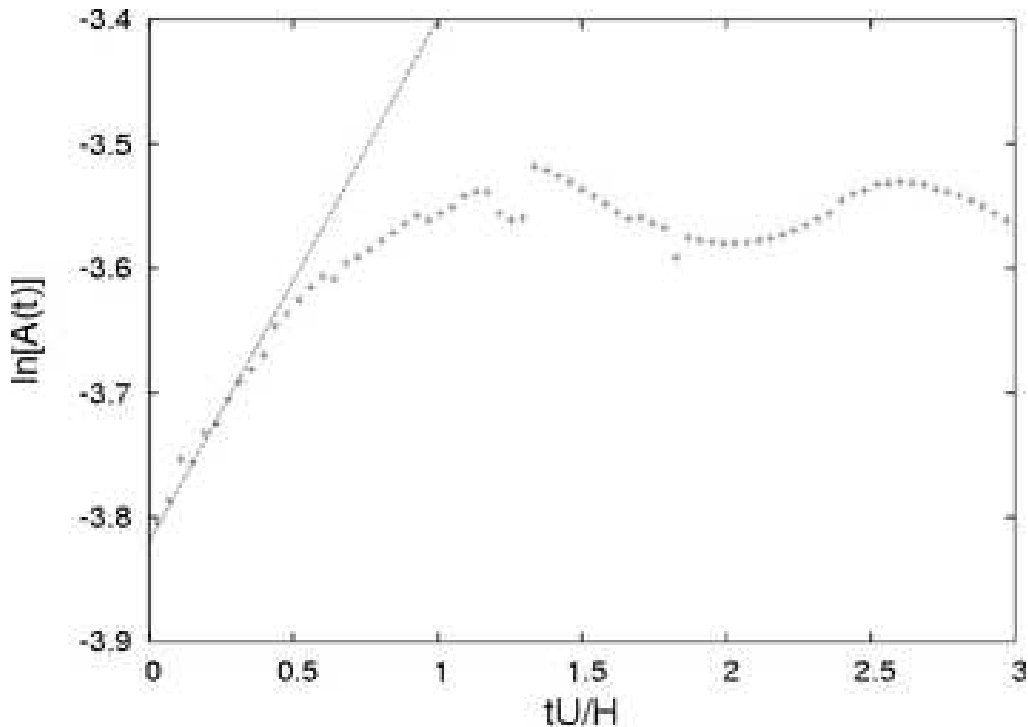}
\caption{Maximum amplitude $A(t)$ versus time (in $H/U$-unit) on a
  $\ln$-linear scale. Solid line represents theoretical linear
  growth and crosses represent simulation.}  
\label{nonlin}
\end{figure}

Some of these results could be expected on physical basis.
In fact, as a first approximation, because of viscous dissipation
effects, the flow can be viewed as a two-layer-flow of two different
viscosity fluids with the less viscous one flowing near the wall. The
simulations perfomed confirm this limit, in fact when viscous heating
form a consistent layer of less viscous liquid near the wall, the
behaviour of this flow tends to be similar to that of a two-layer flow
with the more viscous fluid in the central part and the less
viscous fluid near the wall. This arrangement is common in
transporting heavy viscous oils which are lubricated using a sheath of
lubricating water \citep{josbai97,liren99}. Experiments and
simulations of this two-layer flow type of fluids with high viscosity
ratio predict spatially periodic waves called bamboo waves because of
their shape, and the formation of vortices in the region near the wall
distributed in the trough of the waves \citep{josbai97}. \\
In our simulations, these features can be seen from figure~\ref{temp}
and from figure~\ref{zoom} where a zoom of the flow fields near the
channel wall is shown. In fact following the flow isolines, a
spatially periodic wave can be easily discerned and relatively large
vortices, settled in the middle of the wave troughs, are also
evident.\\    
Moreover, similarly to the core-annular flows with high viscosity
ratio \citep{liren99}, the formation of a mixed profile (with a
counter-flow zone) near the wall, leads to the appearance of vortices
(figure~\ref{vvad}). \\ 
Finally, using the dimensionless numbers  reported in
Table~\ref{simb}, we perfomed a linear stability analysis of the base
profiles given by the lubrication approximation (\ref{eqstaz}) at a 
distance 2/3 of the tube length ($\xi_1^*\simeq 22$) from the
inlet. These analysis indicate that the base flow is already unstable
for $Na\lesssim 120$ even at $Re=120$ and, as it is shown in
figure~\ref{stabmode}, the most dangerous mode for this flow has wave
number $\alpha\approx 7$, corresponding to a wave length
$\lambda\simeq 1.1$ (in $H$-unit) which appears in agreement with that
given by DNS. In fact, as it is shown in figure~\ref{zoom}, at the 
distance $\xi_1^*\approx 22$ from the inlet, a wave length of
$\lambda\approx 1.2$ can be estimated. \\
In order to obtain a closer comparison of the linear stability theory
with the nonlinear results obtained by the numerical code above 
presented, we simulated the case of a channel flow like that
previously described, with conditions at the inlet given by the
solutions of the equations~(\ref{eqstaz}) for $Pe=7400$, $Re=119.4$
and $Na=200$ at $\xi_1=21$, free flow conditions in the remaining
boundary and no-slip conditions at the walls; as initial field inside
the tube we imposed $u_i=\theta=0$. The linear theory predicted for
these profiles a growth rate $\sigma=\alpha c_i\simeq 0.42$ (in
$U/H$-unit). To compare the linear with the nonlinear regimes, the
evolution of the maximum amplitude $A(t)$ with time (in $H/U$-unit) is
plotted in figure~\ref{nonlin}. The maximum amplitude growth shown in 
figure~\ref{nonlin} is given by the evolution of streamlines around a 
short distance from the inlet ($0\le\xi_1\le 1.3$). As initial
amplitude we considered the value of $A(t)$ at the time at which small
perturbations due to round-off errors begin to grow. 
Figure~\ref{nonlin} shows that the initial evolution of the
perturbation is close to that predicted by the linear theory, and then
rapidly starts to deviate as amplitude increases. After only less than
about one time $H/U$ the linear growth completely fails in the
prediction of the amplitude evolution.      
\par
These and other preliminary results of the investigations on viscous
heating effects \citep[e.g.\ ][]{cosmac2003} may help in the
understanding of some common phenomena that may occur during
lava and magma flows. For instance, effects of viscous dissipation can 
efficiently enhance thermal and mechanical wall erosion, and can help  
to understand the reasons of the inadequacy of simple conductive
cooling models commonly used to describe lava and magma
flows. Moreover in volcanic conduits viscous heating could play an
important role on the dynamics of both effusive and explosive
eruptions, influencing directly or indirectly magma gas exsolution and
fragmentation \citep{vedmel2005}. Since magma flowing in conduits and
channels is much hotter than the wall rock, another dimensionless
number ${\mathcal B}=\beta(T_{in}-T_w)$, that compares the imposed
difference of temperature with $\beta$ should be considered ($T_{in}$
and $T_w$ represent inlet and wall temperatures
respectively). However, previous preliminary studies indicate that in
magma flows, viscous dissipation effects can overcomes the thermal
cooling from the walls \citep{cosmac2003}. Moreover, although by
increasing ${\mathcal B}$ the peak in the temperature profile moves
towards the centre of the channel, because of the low magma thermal
conductivities, the flow behaviour is not much different from the case
with ${\mathcal B}=0$ \citep{cosmac2003,sch76}.\\   
\subsection{Validity and limits of the model}
We have seen that when viscous heating is relevant, a special class of
secondary flows can develop in fluids with temperature dependent
viscosities even at low Reynolds numbers. This kind of vortical
structures is locally confined near the walls where there is a large
viscosity gradient and the viscosity is lower.
\par
The results obtained are valid in the limit of a 2D model based on the
full solution of the Navier-Stokes equations although turbulence is
generally three-dimensional even starting with two-dimensional
initial conditions. On the other hand, it is known that the growth of
three-dimensional instabilities may be suppressed by a strong
anisotropy \citep{sommor82,mesmor2001}. This anisotropy can be due to
the presence of a magnetic field \citep{sommor82}, a strong rotation
and/or a density stratification \citep{lil72,hop87,hei93}.\\
As in the isothermal case, the Squire's theorem suggests that for the
linear stability analysis it is sufficient to consider two dimensional
disturbances.  Although the case of core-annular flows with high
viscosity ratio suggests that a 2D model is able to describe well the
flow features observed during the experiments
\citep{josbai97,liren99}, because of the complexity of these
non-isothermal flows, the effects of 3D disturbances on a quasi-2D
flow should be also investigated in order to understand whether the
evolution of three dimensional motions could be able to obscure the
vortical structures described above. 
In our case, we suppose that the strong viscosity stratification
induced by viscous heating could inhibit 3D motions and the 2D model
we used should be able to account for the essential physical
properties of the real systems; however only an extended 3D simulation
can completely confirm this. \\ 
Finally, we note that the numerical scheme we used has a first-order
upwind and it needs very restrictive conditions and a large
computational time in order to be accurate. A more efficient scheme
should be used to permit a more complete parametric study.
\section*{Conclusions}
The thermo-fluid-dynamics of a fluid with strongly
temperature-dependent viscosity in a regime  with low Reynolds
numbers, high P\'eclet and high Nahme numbers were investigated by 
direct numerical simulation (DNS) and the linear stability equations
of the steady thermally developing base flow was studied.\\ 
Our results show that viscous heating can drastically change the flow
features and fluid properties. The temperature rise due to the viscous
heating and the strong coupling between viscosity and temperature can
trigger an instability in the velocity field, which cannot be
predicted by simple isothermal Newtonian models.  \\
Assuming steady thermally developing flow profiles we performed a
linear stabilty analysis showing the important destabilizing effects
of viscous heating.\\   
By using DNS, we showed as viscous heating can be responsible for
triggering and sustaining a particular class of secondary
rotational flows which appear organized in coherent structures similar
to roller vortices. \\ 
We wish our preliminary results can stimulate further more
accurate studies on this intriguing topic, contributing to a more
quantitative comprehension of this problem which has many practical
implications such as in the thermo-dynamics of magma flows in
conduits and lava flows in channels.  

\begin{acknowledgments}
We would like to acknowledge the anonymous referees who strongly
improved the quality of the paper with their useful comments. We also 
thank S. Mandica for his corrections and suggestions.
\end{acknowledgments}
%

%
\bibliography{references}
\end{document}